\documentclass[prl,twocolumn,showpacs,preprintnumbers,amsmath,amssymb,floatfix]{revtex4-1}

\usepackage[dvips]{graphicx}

\usepackage{color}

\begin{document}

\title{Bulk superconducting phase with a full energy gap in the doped 
topological insulator Cu$_x$Bi$_2$Se$_3$}

\author{M.~Kriener, Kouji~Segawa, Zhi~Ren, Satoshi~Sasaki, and Yoichi~Ando} 
\affiliation{Institute of Scientific and Industrial Research, 
Osaka University, Osaka 567-0047, Japan}

\date{\today}

\begin{abstract}

The superconductivity recently found in the doped topological insulator
Cu$_x$Bi$_2$Se$_3$ offers a great opportunity to search for a
topological superconductor. We have successfully prepared a
single-crystal sample with a large shielding fraction and measured the
specific-heat anomaly associated with the superconductivity. The
temperature dependence of the specific heat suggests a fully-gapped,
strong-coupling superconducting state, but the BCS theory is not in full
agreement with the data, which hints at a possible unconventional
pairing in Cu$_x$Bi$_2$Se$_3$. Also, the evaluated effective mass of 2.6$m_e$
($m_e$ is the free electron mass) points to a large mass enhancement in
this material.

\end{abstract}


\maketitle

In the past two years, the three-dimensional (3D) topological insulator
(TI) is attracting a lot of interest as a new state of matter
\cite{Kane,Moore,Zhang}. It is characterized by the existence of a
gapless surface state that emerges because of the non-trivial Z$_2$
topology of the insulating bulk state and is protected
against backscattering by time-reversal symmetry.
The discovery of the 3D TI stimulated the search for a superconducting
(SC) analogue, a time-reversal-invariant {\it topological superconductor}
\cite{fu08a,schnyder08a,qi09b,qi10b,Linder,Sato}, which is characterized
by a fully-gapped, odd-parity pairing state that leads to the emergence
of gapless Majorana surface states. Such a SC phase has
implications on topological quantum computing
\cite{fu08a,fu09a,akhmerov09a,tanaka09a} because of the non-Abelian
Majorana bound state expected to appear in the vortex core.
However, a concrete example of such a topological superconductor is
currently unknown.

In this context, the superconductivity recently found \cite{hor10a} in
Cu$_x$Bi$_2$Se$_3$ is very interesting. Bi$_2$Se$_3$ is a
``second-generation" TI that has a relatively large ($\sim$0.3 eV) band
gap and a simple surface-state structure \cite{zhang09b,xia09a}.
Surprisingly, when Cu is introduced to this system with the nominal
formula Cu$_x$Bi$_2$Se$_3$, superconductivity with a maximum
transition temperature $T_{\rm c}$ of 3.8 K was observed for the doping
range $0.10 \leq x \leq 0.30$, even though the bulk carrier density $n$ 
was only $\sim 10^{20}$ cm$^{-3}$ \cite{hor10a}. 
Note that this $T_{\rm c}$ is uncharacteristically
high for such a low $n$ \cite{Cohen}. Furthermore, the
topological surface state was found to be well-separated from the doped
bulk conduction band in Cu$_x$Bi$_2$Se$_3$ \cite{Wray}. So far, apparent
SC shielding fractions of only up to 20\% have been achieved and the
resistivity always remained finite \cite{hor10a,Wray}, leaving some
doubt about the bulk nature of the superconductivity in this system.
Nonetheless, if this superconductivity is indeed a bulk property of
carrier-doped Bi$_2$Se$_3$, it has a profound implication on the search
of topological superconductors, because (i) it is a potential candidate
\cite{Fu-Berg} to realize a 3D topological superconductor, and (ii) if
its bulk turns out to be an ordinary $s$-wave superconductor, the
topological surface state may turn into a 2D topological superconductor
as a result of a SC proximity effect \cite{fu08a}. Therefore, it is
important to confirm whether the superconductivity is really occurring
in the bulk of Cu$_x$Bi$_2$Se$_3$ and, if so, to elucidate the
fundamental nature of its SC state.

Bi$_2$Se$_3$ has a layered crystal structure (R$\bar{3}$m, space group
166) consisting of stacked Se-Bi-Se-Bi-Se quintuples that are only
weakly van-der-Waals bonded to each other. We call the rhombohedral
[111] direction the $c$ axis and the (111) plane the $ab$ plane. When Cu
is introduced into Bi$_2$Se$_3$, it may either intercalate as Cu$^{1+}$
into the van-der-Waals gaps and act as a donor, or replace Bi as a
substitutional impurity and act as an acceptor \cite{acceptor}; hence,
Cu is an ambipolar dopant \cite{caywood70a,vasko74a}. The nominal
formula of Cu$_x$Bi$_2$Se$_3$ suggests that most Cu atoms in this SC
material occupy the intercalation sites; however, the reported carrier
density of $\sim$10$^{20}$ cm$^{-3}$ \cite{hor10a} corresponds to only
$\sim$1\% of electron doping, which is much smaller than that expected
from the $x$ value. This discrepancy suggests either that most of the
intercalated Cu ions remain inactive as donors, or that substitution of
Bi with Cu also occurs in this material and it almost compensates the
electrons doped by the intercalated Cu. Partly related to such an
uncontrollability of the Cu atoms in Cu$_x$Bi$_2$Se$_3$, the quality of
the SC samples has been poor as mentioned above, and improvements in the
sample quality are indispensable for a solid understanding of the SC
state in this material.

In this Letter, we report a comprehensive study of the basic SC
properties of a Cu-intercalated Bi$_2$Se$_3$ single crystal by means of
resistivity, magnetization, and specific heat measurements. For the
first time in this material, we observed zero-resistivity and a
specific-heat jump at the SC transition. The apparent shielding fraction
of our sample exceeds 40\%, and the specific-heat data confirms the bulk
nature of the superconductivity. Most importantly, the temperature
dependence of the specific heat suggests a fully-gapped, strong-coupling
SC state, but the data do not fully agree with the strong-coupling BCS
calculation. This suggests that the pairing symmetry may not be simple
isotropic $s$-wave. Furthermore, the effective mass is found to be
2.6$m_{\rm e}$ ($m_{\rm e}$ is the free electron mass), suggesting a
change in the bulk band curvature.


Single crystals of Bi$_2$Se$_3$ were grown by melting stoichiometric
amounts of elemental shots of Bi (99.9999\%) and Se (99.999\%) in sealed
evacuated quartz glass tubes at 800$^{\circ}$C for 48 h, followed by a
slow cooling to 550$^{\circ}$C over 48 h and keeping at that temperature
for 24 h. The crystals were cleaved and cut into rectangular pieces, and
then the Cu intercalation was done by an electrochemical technique under
inert atmosphere inside a glove box, using CuI reagent in CH$_3$CN
solvent. The sample was wound with a Cu wire, and a Cu stick was used as
counter and reference electrode. The current was fixed at typically
10 $\mu$A. The concentration of intercalated Cu was determined from the
weight change before and after the intercalation process, and the sample
was briefly annealed afterward. The particular sample used in the
present study was 3.9$\times$1.6 mm$^2$ in the $ab$ plane with a
thickness of 0.40 mm, and its Cu concentration was $x=0.29$. In fact, by
employing electrochemical intercalation, we found that samples with $x$
of up to $\sim$0.5 become superconducting; the precise phase diagram is
currently under investigation.

The magnetization data were measured with a commercial SQUID
magnetometer (Quantum Design, MPMS). The resistivity $\rho_{xx}$ and the
Hall resistivity $\rho_{yx}$ were measured by a standard six-probe
technique, where the electrical current was applied in the $ab$ plane.
The specific heat $c_p$ data were taken by a relaxation-time method
using a commercial system (Quantum Design, PPMS); the addenda signal was
measured before mounting the sample and was duly subtracted from the
measured signal. The $c_p$ measurement was done in zero-field (for the
SC state) and in 2 T applied along the $c$ axis (for the normal state),
and the change of the addenda signal between the two were found to be
negligible.

\begin{figure}
\includegraphics[width=8.5cm,clip]{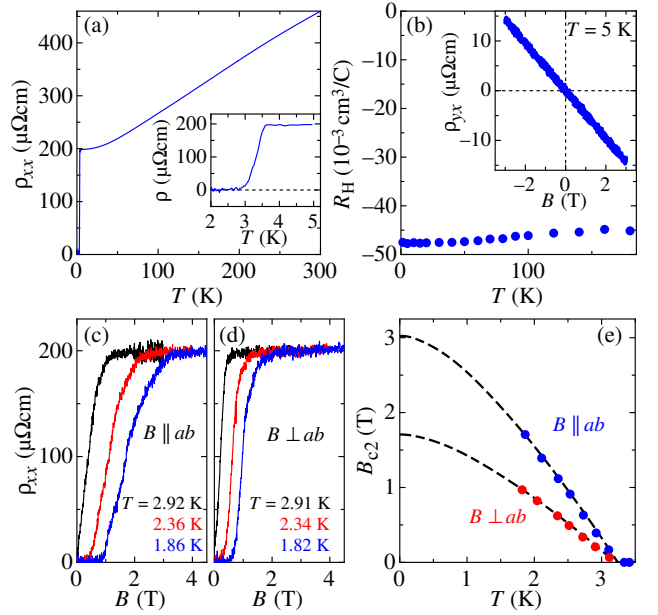}
\caption{(color online) 
(a) $\rho_{xx}(T)$ data of the Cu$_{0.29}$Bi$_2$Se$_3$ sample.
(b) Temperature dependence of $R_{\rm H}$; inset shows the $\rho_{yx}(B)$
data at 5 K.
(c,d) $\rho_{xx}(B)$ data for $B \parallel ab$ and $B \perp ab$, 
respectively. 
(e) $B_{\rm c2}$ vs. $T$ phase diagram determined from the midpoint in 
$\rho_{xx}(B)$ at various temperatures; dashed lines show the WHH behavior.
The midpoint definition for $B_{\rm c2}$ gives $T_{\rm c}$ = 3.2 K 
consistent with the $M(T)$ data. }
\label{rhoplot}
\end{figure}

Figure 1 shows the results of the transport measurements. In zero field,
the onset of the SC transition occurs at 3.6 K and zero-resistivity was
observed at 2.8 K [Fig. 1(a)]. From the magnetic-field ($B$) dependence
of $\rho_{xx}$ [Figs. 1(c) and (d)], the upper critical field $B_{\rm
c2}(T)$ defined as the midpoint of the resistivity transition
\cite{midpoint} is obtained for the two principal field directions [Fig.
1(e)], and the extrapolation to 0 K using the
Werthamer-Helfand-Hohenberg (WHH) theory gives $B_{c2,\perp}(0)$ and
$B_{c2,\parallel}(0)$ (perpendicular and parallel to the $ab$ plane) of
1.71 and 3.02 T, respectively [Fig. 1(e)]. As shown in the inset of Fig.
1(b), $\rho_{yx}$ is completely linear in $B$, suggesting the dominance
of only one type of bulk carriers. The Hall coefficient $R_H$ was found
to be only weakly temperature dependent [Fig. 1(b)] and gives the
electron density $n$ = 1.3$\times$10$^{20}$ cm$^{-3}$.

\begin{figure}
\centering
\includegraphics[width=8.5cm,clip]{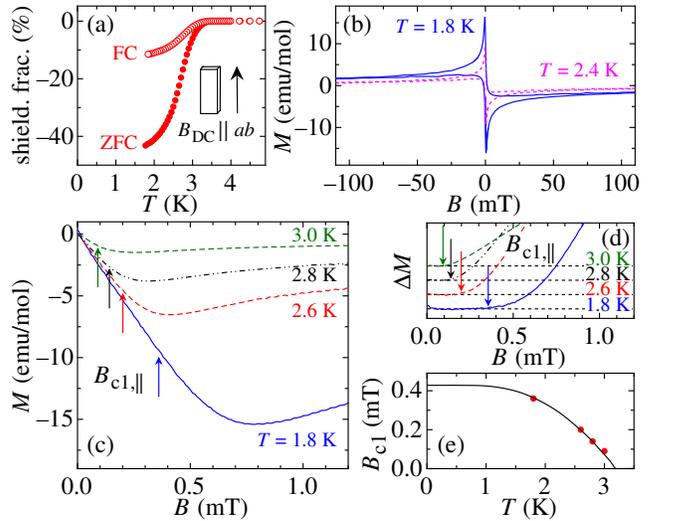}
\caption{(color online) 
(a) Temperature dependence of the apparent shielding fraction of
Cu$_{0.29}$Bi$_2$Se$_3$ measured in $B = 0.2$ mT $\parallel ab$.
(b) $M(B)$ curves at 1.8 and 2.4 K after subtracting the diamagnetic 
background. 
(c) Initial $M(B)$ behavior after ZFC to various temperatures.
Arrows mark the position of $B_{c1,\parallel}$. Note the very small 
magnetic-field scale.
(d) Plots of $\Delta M \equiv M - aB$, where $a$ is the initial slope, 
and the determination of $B_{c1,\parallel}$ shown by arrows.
(e) $B_{c1,\parallel}$ vs.\ $T$ phase diagram; the solid line is a fit within 
the local dirty limit.} 
\label{chiplot}
\end{figure}

Figure 2 summarizes the magnetization $M$ measurements.
To minimize the effect of the demagnetization factor, those measurements
were made for $B \parallel ab$ [see the sketch in Fig. 2(a)]. The
temperature dependence of the diamagnetic shielding fraction for the
zero-field-cooled (ZFC) and field-cooled (FC) measurements are shown in
Fig. 2(a), where the onset of the Meissner signal occurs at $T_{\rm c}$
= 3.2 K and the apparent shielding fraction reaches 43\% at 1.8 K
\cite{fraction}. Note that this Meissner $T_{\rm c}$ corresponds to the
midpoint of the resistivity transition. Neither ZFC nor FC data saturate
at 1.8 K. 

Magnetization $M(B)$ curves are shown in Figs. 2(b) and (c); each data set was
obtained after cooling to its respective temperature from above $T_{\rm c}$ in
zero field, and the background diamagnetism, which can be easily
determined at $B > B_{c2,\parallel}$, is subtracted from the data. As
already noted by Hor {\it et al.} \cite{hor10a}, the lower critical
field $B_{\rm c1}$ is very small: Using the deviation of the $M(B)$ curve
from its initial linear behavior as a measure of $B_{c1,\parallel}$
[Fig. 2(d)], we obtained the $B_{c1,\parallel}(T)$ data shown in Fig.
2(e). To determine the 0-K limit, we used $B_{\rm c1}\propto 1/\lambda_{\rm
eff}^2 \propto [(\Delta(T)/\Delta(0)) \tanh(\Delta(T)/2k_{\rm B} T)]$ for the
local dirty limit \cite{note_Bc1} to fit the extracted data points
($\lambda_{\rm eff}$ is the effective penetration depth and $\Delta$ is
the SC gap \cite{note_Delta}), and obtained $B_{c1,\parallel}^{\rm
app}(0)$ = 0.43 mT. 
For the quantitative analysis discussed later, this apparent
value was corrected for the demagnetization effect, though it
is small for $B \parallel ab$: Using the approximation given
for the slab geometry \cite{brandt99a}, we obtain $B_{c1,\parallel}(0)$
= $B_{c1,\parallel}^{\rm app}(0) / {\rm tanh}\sqrt{0.36b/a}$ = 0.45 mT,
where $b/a = 3.9/0.40$ in our case. Note that the flux pinning in the
present system is weak as evidenced by the low irreversibility field of
$\sim$0.1 T at 1.8 K [Fig. 2(b)].

\begin{figure}
\centering
\includegraphics[width=8.5cm,clip]{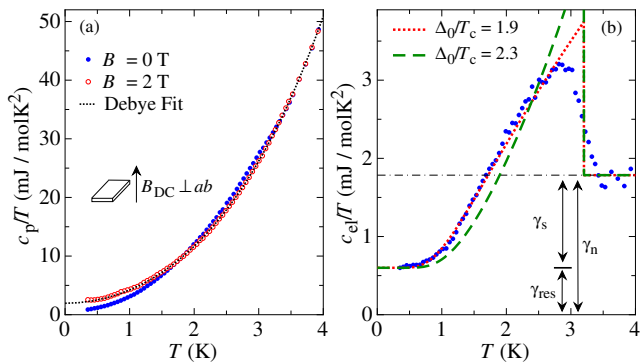}
\caption{(color online) 
(a) $c_p(T)/T$ data measured in 0 and 2 T applied along the $c$ axis;
the latter represent the normal-state behavior. The dashed line is a fit
to the 2-T data using the standard Debye formula. (b) Electronic term
$c_{\rm el}/T$ in 0 T obtained after subtracting the phonon term
determined in 2 T. The dotted line shows the calculated $c_{\rm el}/T$
curve given by strong-coupling BCS theory with $\alpha$ = 1.9; the
dashed line is the BCS curve for $\alpha$ = 2.3, which is obtained from
$B_{\rm c}$, $N_0$, and $T_{\rm c}$. The horizontal dash-dotted
line denotes the value of $\gamma_{\rm n}$, and its breakdown to
$\gamma_{\rm s}$ and $\gamma_{\rm res}$ is indicated.}
\label{cpplot} 
\end{figure}

The temperature dependence of $c_p$ is shown in Fig. 3(a) as $c_p/T$
vs.\ $T$ for the SC state ($B$ = 0 T) and the normal state achieved by
applying $B \perp ab$ of 2 T $(> B_{c2,\perp})$. 
As shown by the dotted line
in Fig. 3(a), a conventional Debye fit to the normal-state data below 4
K using $c_p = c_{\rm el} + c_{\rm ph} = \gamma_{\rm n} T + A_3 T^3 +
A_5 T^5$, with the normal-state specific-heat coefficient $\gamma_{\rm
n}$ and the coefficients of the phononic contribution $A_3$ and $A_5$,
yields a good description of the data. The obtained parameters are
$\gamma_{\rm n}$ =1.95 mJ/molK$^2$, $A_3$ = 2.22 mJ/molK$^4$
\cite{Debye}, and $A_5$ = 0.05 mJ/molK$^6$. Subtracting the phononic
contribution from the zero-field data gives the electronic specific heat
$c_{\rm el}$ in the SC state plotted in Fig. 3(b), revealing a clear
jump around $T_{\rm c}$. This provides compelling evidence for bulk
superconductivity in Cu$_x$Bi$_2$Se$_3$. In passing, we note that our
$c_p$ data in 2 T do not exhibit any Schottky anomaly related to
electron spins, suggesting that there is no local moment possibly
associated with Cu$^{2+}$ ions.

From the above results, one can estimate various basic parameters.
Assuming a single spherical Fermi surface, one obtains the Fermi wave
number $k_{\rm F}$ = $(3\pi^2 n)^{1/3}$ = 1.6 nm$^{-1}$. The effective
mass $m^*$ is evaluated as $m^*=(3\hbar^2\gamma_{\rm n})/(V_{\rm mol}
k_{\rm B}^2 k_{\rm F})$ = 2.6$m_{\rm e}$, with the molar volume of
Bi$_2$Se$_3$ $V_{\rm mol} \approx 85$ cm$^3$/mol. Note that the
effective mass of pristine Bi$_2$Se$_3$ is $\sim$0.2$m_{\rm e}$
\cite{Kohler}, so there is an order-of-magnitude mass enhancement in
Cu$_x$Bi$_2$Se$_3$ \cite{CBM}. Since electron correlations are weak in
Bi$_2$Se$_3$, the origin of this enhancement is most likely a 
change in the band curvature near the Fermi level. 
From $B_{c2,\perp}$ = 1.71 T, the coherence length
$\xi_{ab} = \sqrt{\Phi_0/(2\pi B_{c2,\perp})}$ = 13.9 nm is obtained,
while from $B_{c2,\parallel}$ we use $\xi_{ab}\xi_c = \Phi_0/(2\pi
B_{c2,\parallel})$ and obtain $\xi_c$ = 7.9 nm. Since we have the
$B_{\rm c1}$ value only for $B \parallel ab$, we define the effective GL
parameter $\kappa_{ab} \equiv
\sqrt{\lambda_{ab}\lambda_c/\xi_{ab}\xi_c}$ and use $B_{c1,\parallel} =
\Phi_0 \ln{\kappa_{ab}} / (4\pi \lambda_{ab}\lambda_c)$ together with
$B_{c2,\parallel}/B_{c1,\parallel} = 2\kappa_{ab}^2/\ln{\kappa_{ab}}$
\cite{Clem} to obtain $\kappa_{ab} \approx$ 128. We then obtain the
thermodynamic critical field $B_{\rm c} = \sqrt{B_{c1,\parallel}
B_{c2,\parallel} / \ln{\kappa_{ab}}}$ = 16.7 mT. 

To analyze $c_{\rm el}/T$ in the SC state shown in Fig. 3(b), we tried
to fit the BCS-type temperature dependence to the data. Since the simple
weak-coupling BCS model does not describe the $c_{\rm el}/T$ data (not
shown), we use the modified BCS model applicable to strong-coupling
superconductors as proposed in Ref. \cite{padamsee73a}, where it is
called ``$\alpha$ model" with $\alpha = \Delta_0/T_{\rm c}$ and
$\Delta_0$ is the SC gap size at 0 K. We note that strong coupling means
$\alpha >\alpha_{\rm BCS}=1.764$, and that this model still assumes a
fully-gapped isotropic $s$-wave pairing. Using the theoretical curve
$c_{\rm el}^{\rm BCS}$ of the $\alpha$ model \cite{padamsee73a}, we
tried to reproduce the experimental data with $c_{\rm el}(T)/T =
\gamma_{\rm res} + c_{\rm el}^{\rm BCS}(T)/T$. 
Note that the parameter $\gamma_{\rm res}$ is necessary for describing 
the contribution of the non-SC part of the sample \cite{NSC}; 
also, the theoretical term $c_{\rm el}^{\rm
BCS}/T$ is set to yield $\gamma_{\rm s}$ (= $\gamma_{\rm n} -
\gamma_{\rm res}$) at $T > T_{\rm c}$. 

It turned out that with $\alpha$ = 1.9, $\gamma_{\rm res}$ = 0.6
mJ/molK$^2$, and $\gamma_{\rm s}$ = 1.185 mJ/molK$^2$, the experimental
data is reasonably well reproduced and the entropy balance is satisfied,
as shown in Fig. 3(b) by the dotted line and the dash-dotted line \cite{c_el}. This result
strongly suggests that the SC state of Cu$_x$Bi$_2$Se$_3$ is fully
gapped. The resulting $\gamma_{\rm n}$ (= $\gamma_{\rm res} +
\gamma_{s}$) value of 1.785 mJ/molK$^2$ slightly deviates from the
$\gamma_{\rm n}$ value estimated from the Debye fit to the normal-state
data in 2 T, 1.95 mJ/molK$^2$. This slight difference ($\sim$9\%) might
be the result of a possible field dependence of the normal-state
Sommerfeld parameter, which has to be clarified in future studies. 

To gain further insight into the nature of the SC state in
Cu$_x$Bi$_2$Se$_3$, we examine the implication of the obtained
$\gamma_{\rm s}$ value: The density of states (DOS) $N_0$ is calculated from
$\gamma_{\rm s}$ via $N_0 = \gamma_{\rm s}/(\pi^2 k_{\rm B}^2/3)$ = 1.51
states/eV per unit cell, which is large for a low-carrier-density system
and is in accord with the ``high" $T_c$. 
This value allows us to calculate $\Delta_0$
through the expression for the SC condensation energy $\frac{1}{2}N_0
\Delta_0^2= (1/2\mu_0) B_{\rm c}^2$. With $B_{\rm c} \approx$ 16.7 mT
already calculated, we obtain $\Delta_0$ = 7.3 K which gives the
coupling strength $\alpha$ = $\Delta_0/T_{\rm c}$ = 2.3. This exceeds
the BCS value of 1.764 and hence Cu$_x$Bi$_2$Se$_3$ is a strong-coupling
superconductor, as was already inferred in our analysis of the $c_{\rm
el}/T$ data. More importantly, the $\alpha$ value of 2.3 obtained from
$\gamma_{\rm s}$ is too large to explain the $c_{\rm el}(T)$ data within
the strong-coupling BCS theory: As shown in Fig. 3(b) with the dashed
line, the expected BCS curve for $\alpha$ = 2.3 does not agree with the
data at all. This probably means that the actual temperature dependence
of $\Delta$ in Cu$_x$Bi$_2$Se$_3$ is different from that of the BCS theory, 
which suggests
that the pairing symmetry may not be the simple isotropic $s$-wave.
Obviously, a direct measurement of $\Delta_0$ and $\Delta(T)$ is
strongly called for. On the other hand, the low-temperature behavior of
$c_{\rm el}(T)$ robustly indicates the absence of nodes and points to a
fully-gapped state. It will be interesting to see if the fully-gapped,
time-reversal-invariant $p$-wave state proposed for Cu$_x$Bi$_2$Se$_3$
\cite{Fu-Berg} would provide a satisfactory explanation of our data.

In summary, we report a comprehensive study of the superconductivity in
Cu$_x$Bi$_2$Se$_3$ by means of resistivity, magnetization, and
specific-heat measurements on a single crystal with $x=0.29$ that shows,
for the first time in this material, zero-resistivity and a shielding
fraction of more than 40\%. An analysis in the framework of a
generalized BCS theory leads to the conclusion that the
superconductivity in this system is fully gapped with a possibly non-BCS
character. The fully-gapped nature qualifies this system as a candidate
for a topological superconductor: Since this system hosts a topological
surface state above $T_{\rm c}$ \cite{Wray}, depending on whether the
parity of the bulk SC state is even or odd, either the surface or the
bulk should realize the topological SC state associated with intriguing
Majorana edge states.

We acknowledge S. Wada for technical assistance. We thank L. Fu, S. Kuwabata, 
K. Miyake, and Y. Tanaka for helpful discussions. This work was supported
by JSPS (KAKENHI 19674002 and Next-Generation World-Leading Researchers Progaram), 
MEXT (Innovative Area ``Topological Quantum Phenomena" KAKENHI 22103004), and 
AFOSR (AOARD 10-4103).


\begin{thebibliography}{10}
\bibitem{fraction}
This shielding fraction becomes 41\% if the demagnetization 
effect \cite{brandt99a} is considered.
\bibitem{note_Delta}
For the calculation of $\lambda_{\rm eff}$, we used the strong-coupling 
$\Delta(T)$ with $\alpha$ = 1.9 consistent with the $c_p$ analysis.
\bibitem{CBM}
The contribution of the second conduction band minima (Ref.
\onlinecite{Kohler}), whose effective mass is only $\sim$0.1$m_{\rm e}$,
does not significantly alter this conclusion.
\bibitem{NSC}
In a homogeneous nodal superconductor with 100\% volume fraction,
impurity scattering leads to a finite DOS even at $T=0$. However, the
sample used here consists of superconducting and non-superconducting
(metallic) parts, and the volume fraction of the latter ($\sim$ 60\%)
entirely accounts for the observed residual DOS, which is 33\% of
the total DOS.
\bibitem{c_el}
We allowed $\gamma_{\rm res}$ and $\gamma_{s}$ to be tuned independently.
\end{thebibliography}
\end{document}